# OF MICE AND MEN: SPARSE STATISTICAL MODELING IN CARDIOVASCULAR GENOMICS[1]

BY DAVID M. SEO, PASCAL J. GOLDSCHMIDT-CLERMONT
AND MIKE WEST

*Duke University, University of Miami and Duke University*

In high-throughput genomics, large-scale designed experiments are becoming common, and analysis approaches based on highly multivariate regression and anova concepts are key tools. Shrinkage models of one form or another can provide comprehensive approaches to the problems of simultaneous inference that involve implicit multiple comparisons over the many, many parameters representing effects of design factors and covariates. We use such approaches here in a study of cardiovascular genomics. The primary experimental context concerns a carefully designed, and rich, gene expression study focused on gene-environment interactions, with the goals of identifying genes implicated in connection with disease states and known risk factors, and in generating expression signatures as proxies for such risk factors. A coupled exploratory analysis investigates cross-species extrapolation of gene expression signatures—how these mouse-model signatures translate to humans. The latter involves exploration of sparse latent factor analysis of human observational data and of how it relates to projected risk signatures derived in the animal models. The study also highlights a range of applied statistical and genomic data analysis issues, including model specification, computational questions and model-based correction of experimental artifacts in DNA microarray data.

**1. Introduction.** As part of a program of research in the genomics of atherosclerosis, we have developed studies that involve evaluation of DNA microarray gene expression data, and other forms of molecular and physiological data, in both experimental and observational contexts. This paper

Received February 2007; revised March 2007.
[1]Supported by NSF Grant DMS-03-42172 and NIH Grants NHLBI 1P01-HL-73042-02 and 5RO1-HL72208-03.
Supplementary material available at http://imstat.org/aoas/supplements

represents a case study of experimental data from animal models coupled with human observational data. The study highlights questions of gene identification related to disease risk factors, assessment of expression-based signatures of physiological and disease states based on animal model designed experiments, and evaluation of such signatures in data arising from human observational studies. The analysis is developed in a class of multivariate regression and factor models utilizing sparsity priors; the application is illustrative of the utility of this approach in high-throughput genomic studies. The paper describes aspects of the modeling framework, including conceptual, practical and computational questions. These are developed throughout the applied studies in two model contexts: multivariate regression and anova applied to the designed experiment on mice models, and multivariate factor analysis of human data. We also highlight the biological relevance and specific methods of extrapolation of expression signatures derived in controlled experimental conditions to observational settings.

**2. Cardiovascular genomics and atherosclerosis.** Atherosclerosis—the development of hardened, flow-limiting plaque within arterial vessels—is a chronic human condition with a complex developmental process that is driven by, and responsive to, many lifestyle and environmental factors overlaying genetic determinants. The gradual build-up of arterial plaque based on blood-borne fats, cholesterols, collagens, cell proliferation, calcium and other substances narrows the inside of the artery and can restrict blood flow. Arterial plaques may also rupture to cause clots with severe and often deadly results. The process of atherogenesis plays roles in coronary artery and cerebrovascular diseases, and is thus central to the leading causes of chronic illness and death in the Western world. Inherent and genetic risk factors for development and aggressiveness of the disease include gender, age and family history of premature cardiovascular disease. Atherosclerosis starts in early ages and progresses throughout our lifetime, to greater or lesser degrees as a function of these and other genetic factors as well as lifestyle activities. Among controllable environmental risk factors, diet is key. Dietary fats are fundamental to the drivers of serum cholesterol distribution and triglyceride levels that are key contributors to disease risk. Improved understanding of the progression of atherosclerosis, the roles of known risk factors and their interactions with genetic factors, are needed to improve risk assessment and therapy. Here we are concerned with questions of profiling disease states and progression using DNA microarray data as part of this overall enterprise.

Some of our prior work has investigated expression data derived from the inner walls of human aorta. Using aortic tissues from heart donors, this led to the identification of signatures predictive of atherosclerotic disease burden. The extent of disease development within the inner walls of the aorta can be assessed visually for signs of lesions linked to advanced plaque development, and also staining methods that are expected to reflect early atherosclerotic plaque development in terms of fatty streaks [11, 12]. Gene signatures based on clustering and binary regression tree model analysis to predict "minimal" versus "advanced" disease using simple binary summaries of staining and lesion were developed in [51]. A challenge in this area is that of more precise measures of disease extent; the traditional SudanIV staining methods [11, 12] and the evaluation of relative extent of lesion development are current gold standards but nevertheless represent very noisy and imprecise clinical phenotypes. More advanced clinical diagnostic methods are unable to predict clinical outcomes for a given patient, and these traditional physiological determinations remain the state-of-the-art clinical phenotypes related to disease burden. The existence of advanced raised lesions within the aorta, compared to no such evidence, was central to [51] and we refer to this clinical binary outcome in our exploratory analysis here.

The starting point of the current paper is new experimental data that explores genetic and environmental risk factors in atherosclerosis in animal models. Before genomic data can be widely used to identify patient risk in human populations, we must first understand which genes and their variants actually reflect and contribute to disease. More specifically, we must understand which genetic variants interact with particular environmental risk factors, and need increased efforts to define controlled studies evaluating responses to environmental exposures. Such studies will naturally continue to heavily require murine (and other) disease models. Some aspects of a new and rich data set generated from such a study on mice have already illuminated the disease process [31]. Some direct comparisons of gene expression data from subsets of mice in the experiment to be described further below identified gene subsets and expression signatures predictive of disease state and linked the results to potential functional associations with the development of arterial repair mechanisms. The current paper develops a broader and detailed study with an expanded experimental mice data set, aiming to explore the cross-connections between environmental risk factors that may modify disease susceptibility by affecting gene expression.

**3. Profiling atherosclerotic disease risk: A gene-environment interaction study.** Our experimental study involves a cross-classified, multi-factorial design to generate gene expression from arterial tissues in the main aortas of mice. The study investigates three key environmental risk factors—Age, Gender and Dietary fat intake—coupled with a key genetic factor related to the ApoE (Apolipoprotein E) gene pathway. ApoE has many known functions linked to fat metabolism and dietary molecular transport [44]. The ApoE protein forms a component of cholesterol complexes that play roles in the transport of triglycerides and in cholesterol distribution among cells. It is further involved in lipid metabolism and the binding of lipoproteins to the LDL receptors that feed into cellular uptake of lipoproteins for cholesterol metabolism [36]. It is well known that ApoE deficiency is a major genetic risk factor for atherosclerosis, *causing* high serum cholesterol and triglyceride levels and leading to premature and advanced disease [34]. The ApoE knockout (ApoE-/-) mouse model closely mimics human atherosclerosis both in the spontaneous appearance of lesions and in the distribution of lesions within blood vessels. In contrast, wild type (WT) mice have intrinsic resistance to atherosclerosis.

TABLE 1
*Design layout, (0/1) coding of factor levels, $\beta_{\cdot,\cdot}$ notation for effects, and actual sample sizes (in parentheses) for the mouse model expression experiment (see also Table 1 of the supplementary material)*

|  | Wild Type (0) | | ApoE-/- (1) | |
|---|---|---|---|---|
|  | Chow diet (0) | Western diet (1) | Chow diet (0) | Western diet (1) |
| 6 week (0) | | | | |
| Female (0): | $\beta_{g,1}$ (5) | $+\beta_{g,5}$ (5) | $+\beta_{g,2}$ (6) | $+\beta_{g,8}$ (4) |
| Male (1): | $+\beta_{g,4}$ (5) | $+\beta_{g,11}$ (5) | $+\beta_{g,7}$ (8) | $+\beta_{g,14}$ (5) |
| 12 week (1) | | | | |
| Female (0): | $\beta_{g,3}$ (5) | $+\beta_{g,10}$ (6) | $+\beta_{g,6}$ (3) | $+\beta_{g,13}$ (9) |
| Male (1): | $+\beta_{g,9}$ (5) | $+\beta_{g,15}$ (5) | $+\beta_{g,12}$ (3) | $+\beta_{g,16}$ (11) |

Our experimental design provides data on aortic gene expression linked to aging, diet, gender, the ApoE deletion and the interactions of these factors. We aim to explore whether there are patterns of gene expression that vary with dietary fat content, with gender, with aging and with the deletion of the ApoE locus, and to define gene subsets underlying any such patterns for further study, interpretation and annotation. Wild type and ApoE-/-(C57BL/6J) mice from the Jackson laboratories were used, with mice pups weaned at three weeks of age and fed either the regular chow diet or the high-fat, Western diet and then grown to either 6 weeks or 12 weeks of age. Aortic tissue was removed at the 6 or 12 week end point (all mouse work was approved by the Duke University IACUC under protocol A288-02-09).

The design and numbers of mice are in Table 1. The design is a saturated cross-classification of ApoE, Age, Gender and Diet, with several replicates in each cell. This enables study of the impact on aortic gene expression of the risk-related genotype (Wild Type versus ApoE-/-), Age ("young" = 6 week old, versus "old" = 12 week old), gender (female versus male) and dietary fat content (standard, low fat "chow" diet versus high-fat "Western" diet), and all levels of interactions. Roughly half the mice are ApoE-/-mutants, roughly half are older, roughly half are males and roughly half are fed the high fat, Western diet. The balance and replication lead to a well-structured design for the evaluation of expression variation related to main effects and potential interactions of genetic and environmental factors.

Disease development was studied by staining the aortic tissue to assess the extent and nature of lesion development [31]. This confirmed intensification of disease burden with age, diet and ApoE-/- and that disease is markedly increased in the older, mutant mice on the high fat diet. We therefore refer to these risk factors as defining disease related states of increasing severity as a function of level of interaction. Each aorta generated mRNA extraction for microarray analysis at the Duke University Microarray Core Facility. This involved standard quality analysis (Agilent Bioanalyzer), then cDNA synthesis and hybridization to the MG-U74Av2 oligonucleotide microarray (Affymetrix). The resulting samples were processed using the RMA [26, 27] delivering measures, on a $\log_2$ scale, of expression intensities for each of the 12488 gene probe sets on each mouse array.

## 4. Multivariate sparse anova model.

4.1. *Basic model form.* Write $x_{g,i}$ for the expression intensity of gene $g = 1:p$ on sample (mouse) $i = 1:n$. For each gene $g$, a linear anova model is then defined by the $n \times k$ design matrix $H$ of 0/1 entries corresponding to the design and parameters in Table 1. With the $k = 16$ $\beta_{g,\cdot}$ parameters defining the $k$-vector of effects $\beta_g = (\beta_{g,1}, \ldots, \beta_{g,k})'$, we have

$$x_{g,i} = \sum_{j=1}^{k} \beta_{g,j} h_{j,i} + \varepsilon_{g,i} = \beta_g' h_i + \varepsilon_{g,i},$$

where the $k$-vector $h_i = (h_{1,i}, \ldots, h_{k,i})'$ is the $i$th row of $H$. Our analysis assumes a normal residual distribution, $\varepsilon_{g,i} \sim N(0, \psi_g)$ independently. This represents the combination of unexplained biological variation in expression of gene $g$, model misspecification and contributions from technical and measurement error that is idiosyncratic to that gene. A major component of variance is technical. With Affymetrix data analyzed on the $\log_2$ RMA scale, prior experience with many data sets indicates technical variation in the range of about 0.1–0.5, with values around 0.2–0.3 being quite typical, suggesting values of $\psi_g$ will typically range across 0.01–0.25 or thereabouts.

On each mouse $i$ we then have the $p$-vector response $x_i$ defined by $x_i' = (x_{1,i}, \ldots, x_{p,i})$ given by

(4.1) $$x_i = Bh_i + \varepsilon_i,$$

where $B$ is the $(p \times k)$-matrix of regression parameters, or design effects, whose $g$th row is $\beta_g'$. The residual $p$-vector $\varepsilon_i = (\varepsilon_{1,i}, \ldots, \varepsilon_{p,i})'$ is distributed as $\varepsilon_i \sim N(0, \Psi)$, independently over $i$, where $\Psi$ is the $p \times p$ diagonal matrix of elements $\psi_g$ ($g = 1:p$). In full matrix form with $p \times n$ response matrix $X =$ $[x_1, \ldots, x_n]$ and $p \times n$ residual matrix $E = [\varepsilon_1, \ldots, \varepsilon_n]$, we have $X = BH' + E$ with $E \sim N(0, \Psi, I_n)$, the matrix normal distribution with left variance $\Psi$ and right variance the $n \times n$ identity.

4.2. *Multivariate analysis and sparsity.* A key biological perspective is that, though some or all of the experimental groups may be associated with significant changes in gene expression for a number of genes, many genes will be unaffected. Further, the complexity represented by higher-order interactions of design factors is unlikely to be realized for many genes. Hence, we expect each $\beta_g$ vector to have several or all zero elements (apart from the first); the "tall and skinny" matrix of regression parameters $B$ will be sparse. The sparsity pattern is unknown and to be estimated—we aim to identify which entries are nonzero with high probability, and to account for the fact that with so many genes we need to guard against false discovery.

Analysis with some form of shrinkage approach is mandated. Shrinkage using point-mass mixture priors—the workhorse for such problems for some years now—extends standard regression variable selection methodologies to these problems of large-scale comparisons. Such approaches have been well-described and quite widely applied in two sample expression studies [50] and more complex settings [7, 15, 28, 29]. We use practical extensions of the standard point-mass, mixture priors from variable selection in regression [10, 17, 43], following developments in studies of sparse multivariate factor models in [53] that have also been extended to a comprehensive class of latent factor regression models [9, 35]. These prior studies have explored sparsity modeling of expression data in biological pathway studies in cancer; the work here parallels that in the context of atherosclerosis.

4.3. *Models of sparsity of design effects.* The sparsity model is a standard variable selection construct that utilizes indicators $z_{g,j} = 0/1$ such that $z_{g,j} = 1$ if any only if $\beta_{g,j} \neq 0$, coupled with priors $\beta_{g,j} \sim N(\cdot|0, \tau_j)$ describing the distribution of any nonzero effects within design factor $j$, that is, within column $j$ of $B$. This applies to $j = 2\!:\!k$ with nondegenerate models for the effect-inclusion indicators $z_{g,j}$, while the intercepts $\beta_{g,1}$ are treated simply via $\beta_{g,1} \sim N(b, \tau_1)$ for some specified prior mean $b$. The generalization of the standard sparsity models described in [9, 35] is as follows: for each factor $j = 2\!:\!k$, the indicators across variables $g$ are exchangeable, with Bernoulli probabilities $\pi_{g,j}$ subject to the hierarchical model

(4.2) $\qquad \pi_{g,j} \sim (1-\rho_j)\delta_0(\pi_{g,j}) + \rho_j Be(\pi_{g,j}|a_j m_j, a_j(1-m_j)),$

where $Be(\cdot|am, a(1-m))$ is a beta distribution with mean $m$ and precision parameter $a > 0$, and the probabilities $\rho_j$ are modeled as $\rho_j \sim Be(\rho_j|sr, s(1-r))$, independently, where $s > 0$ is relatively large and $r$ a small probability. The basis for this model is that: (a) with the $\rho_j$ likely to be small, the model places a high probability on many zero values among the $\pi_{g,j}$, which in this application is apt as it emphasizes the biological reality that many genes should have no probability of responding to an intervention; and (b) with $m_j$ set at relatively high values, the nonzero $\pi_{g,j}$ will tend to be high, so that corresponding $\beta_{g,j}$ are likely to be nonzero.

Key inferential questions concern the identification of genes (if any) that exhibit significant design effects, and then inferences on those likely nonzero effects. We will address this by evaluating sets of posterior probabilities $\pi^*_{g,j} = Pr(z_{g,j} = 1|X) = Pr(\beta_{g,j} \neq 0|X)$ and point estimates such as $\hat{\beta}_{g,j} = E(\beta_{g,j}|X, z_{g,j} = 1)$, along with other posterior quantities.

4.4. *Sparse regression components for assay artifact correction.* In common with all microarray assays, Affymetrix array data can exhibit variation that results from a combination of numerous experimental and assay sources overlaid on the biologically derived variation of interest. The potential sources include batch effects, unpredictable changes and fluctuations in experimental controls (hybridization temperatures, salinity, etc.), assay reagents, technician practices, equipment settings and so forth. Such "assay artifacts" [9, 35] are often benign and impact on only a few genes per sample and perhaps a few samples. They can, however, also be quite substantial, impacting across an entire sample data set and contributing systematic biases to the summary expression measures for multiple genes. The effects can be particularly marked when samples are collected over a longer period of time, whether or not they are processed at the same laboratory. Some form of the gene-sample specific normalization method is needed to address this. We use information from housekeeping (or "maintenance") genes to help with this. Affymetrix microarrays have a number of gene probes designed to show either no expression at all or to maintain approximately constant levels of expression across diverse biological conditions. Such genes can serve as normalization controls, and the above references have demonstrated the ability of these probesets to provide covariate information capable of capturing some aspects of assay artifact that shows up in genes of interest. Use of a few of the dominant principle components among the housekeeping genes has the ability to capture key aspects of nonbiological assay artifact variation that may selectively impact on gene subsets across samples, and we use these principal components as candidate "assay artifact correction factors" in an expanded model for the vector of expression outcomes. Here we used the first 5 (of $n = 90$) principal components of the selected set of 41 "AFFX" housekeeping probesets on the mouse array. Though these controls may describe significant artifactual contributions to variation in some genes on some samples, we recognize that many genes will be quite unaffected on some or all samples; hence, the use of sparsity priors for regression coefficients on these control factors is apt, providing an ability to develop a parsimonious approach to gene-sample specific, model-based artifact correction.

The extension of the basic sparse anova model of equation (4.1) is immediate. We extend the regression vectors $h_i$ to include an additional 5 entries—the values of the assay artifact covariates. The gene specific regression parameter vectors $\beta_g$ are similarly extended with 5 additional potential regression parameters per gene, subject to the same form of sparsity prior as used for the anova design effects.

4.5. *Complete model specification.* To complete the model specification, we use priors on the remaining model parameters as follows. For each of the $\pi_j$, the required hyperparameters discussed in the previous section are specified as $(r,s) = (0.001, 20)$ and $(m_j, a_j) = (0.9, 10)$ for each $j = 2:k$. A diffuse prior is specified for the intercepts $\beta_{g,1} \sim N(b, \tau_1)$ with $(b, \tau_1) = (8, 100)$ (recall we have data on the $\log_2$ Affymetrix scale for which expression values range from near zero to 15–16). For $j > 1$, the $\tau_j$ parameters are variances defining the levels of change in average expression for any genes that do associate with factor group $j$. Thus, $\sqrt{\tau_j}$ is on the $\log_2$ expression scale and realistic prior expectations indicate values outside a range of 0.5–5 are unlikely. We adopt relatively diffuse conditionally conjugate inverse gamma priors $\tau_j^{-1} \sim Ga(\tau_j^{-1}|5/2, 1/2)$ to reflect this. The residual variance terms $\psi_g$ have rather diffuse inverse gamma priors $\psi_g^{-1} \sim Ga(\psi_g^{-1}|25/2, 0.1/2)$. With Affymetrix data on the $\log_2$ RMA expression, or similar, scale, experience with multiple studies of designed experiments suggest standard deviation values in the range 0.1–0.3 or thereabouts, and this prior is consistent with and informed by such experiences.

4.6. *Computations.* Model fitting uses standard Markov chain Monte Carlo (MCMC) methods in a relatively routine (though high-dimensional) Gibbs sampling format with some Metropolis–Hastings components. Details appear in [9, 35] for a more general model framework. Software implementing the analysis is freely available—the BFRM (Bayesian Factor Regression Model) code implements sparse statistical models for high-dimensional data analysis based on latent factor models coupled with sparse regression and anova; the analysis here uses just the latter components. The MCMC was initialized at values consistent with the prior and data and the simulation run for 10,000 iterations to achieve nominal burn-in before saving and summarizing samples for a series of 100,000 iterations.

The MCMC analysis generates posterior samples for all model parameters: $B$, the indicators $z_{g,j}$, the values of the posterior effect probabilities $\pi^*_{g,j}$, the $\tau_j$, the residual, idiosyncratic variances in $\Psi$ and the sparsity base rates $\rho_j$. These samples can be summarized to produce (Monte Carlo approximations to) posterior means and uncertainty evaluations for all quantities.

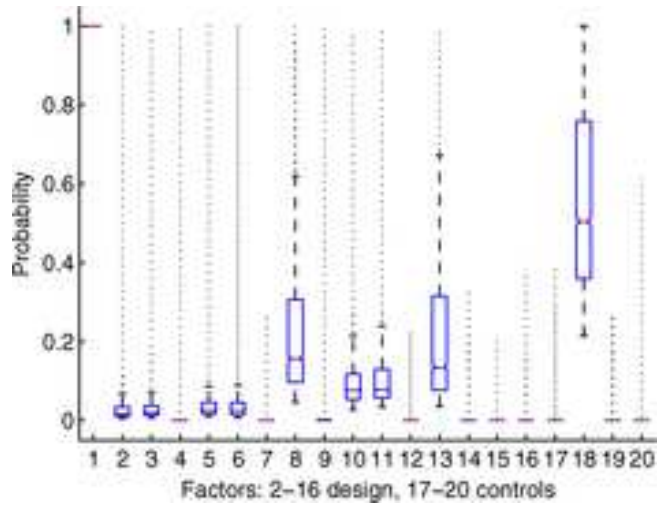

Fig. 1. *Boxplots of the posterior probabilities of nonzero effects $\pi^*_{j,j}$ for all $g = 1:p$ genes within each of the $j = 2$–$16$ design groups ($j = 1$ represents the intercept), together with those for the four included assay artifact control covariate factors $j = 17$–$20$.*

## 5. Analysis of murine atherosclerosis gene-environment study.

5.1. *Analysis set-up.* The analysis was run on RMA expression indices of $p = 5328$ Affymetrix probesets (genes) on the mice arrays. The restriction to these genes was based on the interest in mapping to human samples [discussed further below (Section 7)] and this involved an initial reduction to 7381 gene probesets on the mice array that have homologues on the Affymetrix HU95av2 human array used to assay RNA from human aorta samples in our prior study [51]. Of these 7381, we then removed probesets showing little or no variation above noise levels (absolute change across samples less than 0.25, and median across samples lower than 5.0, applied to both mouse and human samples separately) to deliver the consensus 5328 genes for study. Some investigations of aspects of the resulting posterior distribution over all model parameters are now detailed.

5.2. *Significant gene-design factor associations.* Examination of the probabilities $\pi^*_{g,j}$ of nonzero effects (Figure 1) indicates interest with respect to the ApoE-/-interactions ApoE.Age (design group $j = 6$), ApoE.Diet (group $j = 8$) and ApoE.Age.Diet (group $j = 13$), as expected; these are the key risk factor groups. The extent of the evidence for significant effects for any individual gene (Figure 1 and Table 2) indicate also that, beyond genes in these risk categories, there are small numbers of genes showing evidence of expression changes as a function of Age, Sex, Diet and (very modestly) their second order interactions. These are also expected.

In addition, just one of the assay artifact control factors appears to contribute significantly to expression patterns of a substantial number of genes, while the others are of negligible importance. Of the subsets of genes linked to the environmental factors and their relevance as risk measures within the mouse disease model, Figure 2 shows how the numbers break down into genes that are unique to the selected subsets on each interaction, and those that overlap either two or all three of the interaction subsets, based on a purely nominal threshold of $\pi_{g,j}^* > 0.95$. Genes in the central three-way overlapping group should be of clear biological interest for further study in relation to known or hypothesized mechanisms of disease development and maturation, while the sets in the two-interaction overlap groups bear additional study. These sets of genes are fully described in a table in the web-based supplementary material, and aspects of the biological relevance and interpretation appears in Sections 5.3 and 6 below.

Exploring ranked lists of genes by each factor group provides insights into the strength and nature of associations with risk factors, and potentially on the biological underpinnings based on genes so identified. Selection of genes might include studying posterior probabilities, odds ratios or methods of false discovery control including posterior expected false discovery computations [15, 16], in tandem with estimated effects parameters $\beta_{g,j}$. We explore genes ranked by posterior probabilities and then by absolute values of the approximate posterior means of the $\beta_{g,j}$ effects. The latter are reported along with the gene identifiers for the intersecting risk groups noted above. We do not regard this as anything more than an exploratory analysis to prioritize subsets of genes for exploration in the biological literature or for further study. In particular, we do not regard this exploratory analysis summary as at all linked to a formal decision process, and view much of the statistical literature on false discovery as overly stylized and of little relevance to summarizing a data analysis. Ultimately, the posterior probabilities of the nonzero factor effects (already appropriately "corrected" for the inherent multiplicities through the use of the hierarchical shrinkage model) are the key summaries, and biological evaluation the only proper basis for including/excluding any genes with respect to defined statistical significance. That said, for summary in examples such as this, presentation of analysis summaries in terms of patterns of gene expression and gene names requires a choice; we threshold at $\pi_{g,j} > 0.95$ here for the sake of argument. Genes showing high associations with more relaxed thresholding may of course prove to be of interest and this form of post-analysis study may be repeated.

TABLE 2
Numbers of genes with high posterior association probabilities $\pi_{g,j}^*$ for those effects/parameters for which significant associations are identified in the mice models data analysis

| $j$ | Parameter | \#$\{g : \pi_{g,j}^* > q\}$ | | |
|---|---|---|---|---|
| | | $q = 0.90$ | $q = 0.95$ | $q = 0.99$ |
| 2 | ApoE-/- | 49 | 35 | 27 |
| 3 | 12wk | 68 | 57 | 45 |
| 4 | Male | 6 | 6 | 5 |
| 5 | Western diet | 27 | 19 | 8 |
| 6 | ApoE-/-, 12wk | 49 | 40 | 19 |
| 8 | ApoE-/-, Western diet | 173 | 100 | 34 |
| 9 | 12wk, Male | 5 | 2 | 2 |
| 10 | 12wk, Western diet | 8 | 4 | 0 |
| 11 | Male, Western diet | 45 | 29 | 6 |
| 13 | ApoE-/-, 12wk, Western diet | 328 | 228 | 111 |
| 18 | Artifact control factor 2 | 689 | 417 | 119 |

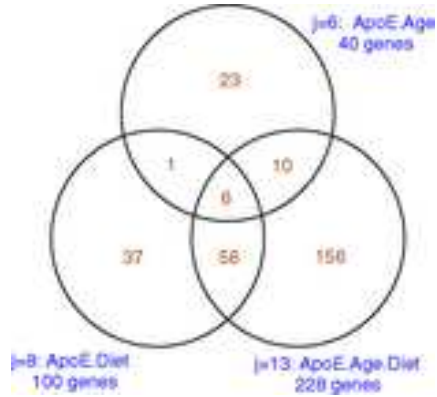

FIG. 2. *Numbers of significant genes ($\pi^*_{g,j} > 0.95$) in each of three key gene-environment interaction design groups, with indication of the numbers of genes in the intersections of the identified gene groups.*

5.3. *Gene-by-factor evaluation.* The expression intensities of four genes are graphed in Figure 3. The data are labeled $x$, and the frames show fitted components of the model for those design factors for which $\pi^*_{g,j} > 0.95$ together with the fitted residuals. This displays the breakdown of expression into significant contributions from these sources for these genes, each of which is known to be related to the atherosclerosis. We use these genes as examples of model-based decomposition and attribution of the patterns of variation in expression to design factors and their interactions.

Osteopontin (named minpontin in the mouse genome), a major player in cardiovascular diseases and in atherosclerosis in particular [18], is involved in mediating processes of cellular adhesion and migration, activities that are central to vascular remodeling in plaque formation and development, among other functions. Here the expression levels of osteopontin are significantly increased with both age and diet risk factors separately among disease prone mice, as well as substantially more so for the $j = 13$ interaction group of very advanced disease, so it plays a role as a generic disease marker. VCAM1 (vascular adhesion molecule 1) is a central marker of the disease, being known to play a pivotal role in the initiation of atherosclerosis and to be highly expressed in arterial endothelial cells in regions predisposed to atherosclerosis and atherosclerotic plaques [13]. VCAM1 is significantly up among older ApoE-/- mice and more so among those on the Western diet. Thrombospondin is well known to be associated with a variety of cellular processes relevant to atherosclerosis, including vascular smooth muscle cell migration, and increased expression levels are markers of advanced disease states in atherosclerosis as in other diseases [37]. The analysis here reflects this, identifying practically large increases in expression only in the advanced disease group—older, Western diet fed ApoE-/- mice—but not due to diet and age alone. TNFR (tumour necrosis nuclear factor receptor) plays a role in the induction of expression of adhesion molecules including VCAM1 and promoting inflammation within vascular endothelial cells [57], and serves as a marker of advanced disease states. We see this reflected in increased expression levels among older ApoE-/- mice on the Western diet, but not across the other design groups. TNFR is also an example of a gene whose expression patterns show significant association with one of the experimental artifact covariates, and this shows how the correction model aids in isolating biologically interpretable components of expression fluctuations from data corrupted with experimental and artifactual noise.

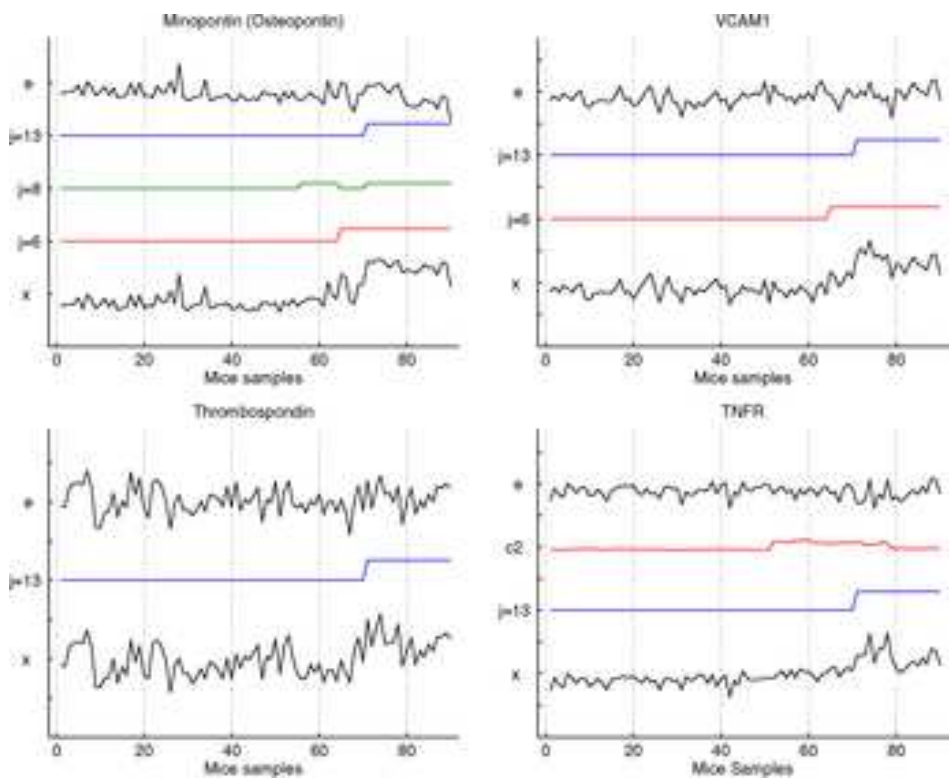

FIG. 3. *Expression decompositions of selected genes in the mice data analysis. In each frame, the plots over time are expression data $x$, fitted model components from design and regression factors showing significant ($\pi^*_{g,j} > 0.95$) contributions for the specific gene, and residuals. Expression is effectively the direct sum of the components and residuals plotted. In each frame the data and components are plotted on the same vertical scales for comparison. Labels $j = 6, 8, 13$ represent the fitted design effects for the three groups, $c2$ the fitted effects of the second assay artifact correction factor, and e the residuals.*

5.4. *Metagene signatures of risk groups.* Several additional figures display summaries of the expression data of these key gene subsets in terms of summary overall measures defining signatures of the subsets. We do this via the factor metagene of any gene set—simply the first principal com-ponent (singular factor) of the subset of genes. This follows the original introduction of the term "metagene" [54, 55] and its use in defining a single numerical "signature" of a set of genes in biological experimental contexts [4, 5, 6, 21, 23], as well as in a number of observational disease studies [20, 22, 41, 45, 51] using gene expression technologies. Given a set of indices $Q \in 1\!:\!p$ with $|Q|=q$, write $X_Q$ for the $q \times n$ matrix of expression values across the $n$ samples. The SVD is $X_Q = ADF$, where $A$ is the $q \times r$ eigenvector matrix with orthonormal columns, $D$ is the $r \times r$ diagonal matrix of singular values, $F$ is the $r \times n$ orthogonal matrix of singular factors, and $r = \min(q, n)$. If $a_{Q,1}$ is the first column of $AD^{-1}$, then $f'_1 = a'_{Q,1} X_Q$ is the first row of $F$ and gives the $n$ values of the first singular factor across the samples. In cases of reduced rank, some elements of $D$ will be zero and $A, D$ and $F$ are reduced by deleting the corresponding columns, diagonal elements and rows, respectively. The vector $a_{Q,1}$ represents a signature of the gene subset $Q$, and the factor $f_1$ the corresponding metagene evaluated over samples. In cases when $Q$ represents a set of coherently related genes with a dominant overall pattern, such as genes whose expression levels tend to be high in one risk category, $a_{Q,1}$ is a characterizing signature of that category.

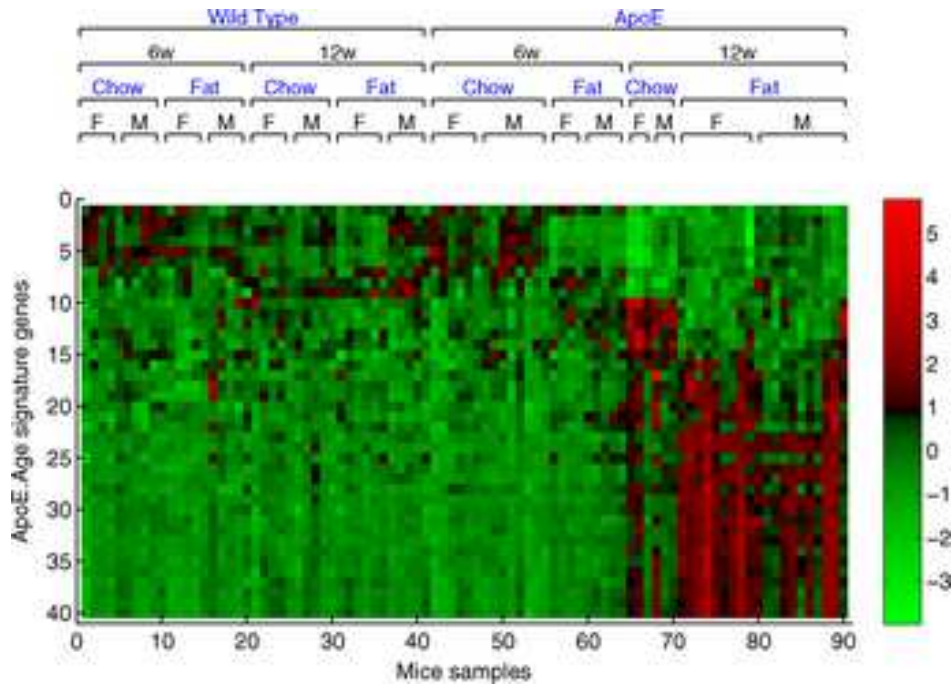

FIG. 4. *Expression image of genes in the mice ApoE.Age signature (design group $j = 6$). The older, ApoE-/- mice are those numbered 65–90 inclusive.*

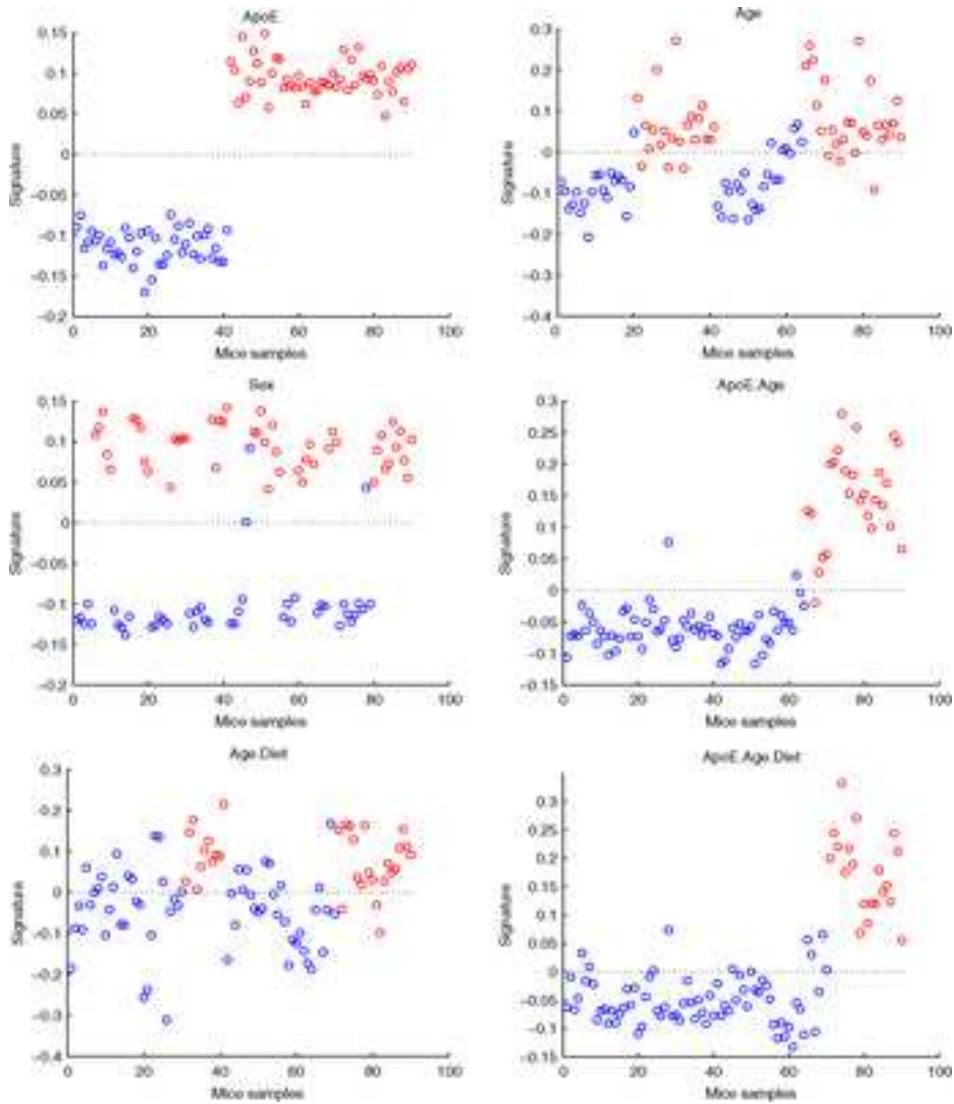

Fig. 5. *Metagene signatures of selected design groups. In each, the graph displays the first principal component of the set of genes showing high probability ($\pi^*_{g,j} > 0.95$) of change in expression with respect to the design group. The color coding simply identifies mice samples in the corresponding group. For example, in the upper-left frame red indicates ApoE-/- while blue indicates wild type; in the centre-right frame, red indicates older ApoE-/- mice; in the lower-left frame, red indicates older mice on the Western diet, and so forth (see also the metagene signature image/heat-map in Figure 1 of the supplementary material).*

In practice, we apply this construction not to the raw data $X_Q$ but to that "corrected" by the fitted model; for each gene, we subtract the contributions from the overall mean and the regression on the artifact control factors (each term of the form $\hat{\beta}_{g,j} \pi^*_{g,j}$, where $\hat{\beta}_{g,j}$ is the posterior mean of $\beta_{g,j}$) to produce centred and artifact-corrected versions of gene expression values.

As one example, Figure 4 shows expression intensities of the $q = 40$ genes characterizing one of the gene-environment interaction groups, the older ApoE-/- mice ($j = 6$). This is an overall, visual "signature" of the group, and the extent of changes in expression among older knockout mice is apparent. The metagene factor generated is plotted across samples in (the center-right frame of) Figure 5; it is clear that this single factor provides a clear and distinctive signature of the risk group. Additional metagenes for other design groups appear in the figure, and highlight the characterization of several of the genetic and environmental factors in terms of single expression signatures of relevant genes.

**6. Genes identified in key risk groups.** The above discussion highlights three key design groups related to gene-environment interactions; genes characterizing these groups are of interest in connection with identifying prognostic summaries of gene expression linked to advancing disease state. The discussion named some such genes that have known functions in the development and progression of atherosclerosis, and similar features arise with multiple other known genes, especially subsets related to the immunological and inflammatory responses characteristic of atherogenesis. We now look at the potential to aid further biological evaluation by perhaps identifying novel candidates linked to the disease process.

The association of age and diet with the development and progression of atherosclerosis is well accepted. One of the results here is the proposal of genes that are associated not just with the presence or absence of atherosclerosis but also at the interaction with age, diet or both. The interaction between atherosclerosis and age is quite clear. The prevalence of atherosclerosis in subjects over the age of 70 exceeds 50%; in men alone, the prevalence approaches 70% by age 70. This is not surprising as mechanisms of aging are also mechanisms involved in atherosclerosis. Two major aging mechanisms are the damage caused by oxidative stress and by the deposition of advanced glycoslyation end products (AGE proteins), both of which lead to extensive apoptosis. Among the list of genes in the ApoE.Age risk group are CD53 and galectin 3, neither of which has, to our knowledge, been directly implicated in the disease process to date.

CD53 is a glycoprotein of the tetraspanin superfamily associated with the recycling enzyme gamma-glutamyl transpeptidase [3]. As such, CD53 plays a role in redox buffering of cells to provide protection from oxidative stress [33, 39]. CD53 expression is elevated in situations of increased oxidative stress such as in rheumatoid arthiritis, radiation damage and also aging, and its expression appears to counter the acceleration of programmed cell death by oxidative damage. Galectin 3 is a member of the beta-galactoside-binding gene family. In cancer, galectin 3 contributes to tumor cell adhesion, proliferation, differentiation, angiogenesis and metastasis [38]. Its primary biological effect appears to be to decrease cell death by making cells less sensitive to pro-apoptotic pathways. In arterial tissues, galectin 3 appears to play a protective role against the damage caused by advanced glycosylation end-products (AGE) [42]. In animal models that carry a galectin 3 knock-out, there is a substantial increase in damage from AGE deposition relative to control animals [24, 25]. The activity of AGE-proteins is a mechanism for diabetes associated damage and also age-related vascular damage [56].

The interaction between atherosclerosis and diet, especially cholesterol, is well established; the mechanism is through deposition of low density lipoprotein cholesterols into the vascular tissues with development of inflammation and apoptosis. Two genes in the ApoE.Diet risk group are Caspase 9 and insulin like growth factor receptor 1 (ILGFR1).

Caspase 9 is part of a group of proteins that play a central role in programmed cell death or apoptosis. Some of the most important mechanisms in the development of atherosclerosis revolve around the pathways triggered by elevated serum cholesterol [32, 40]. One specific mechanism is that high levels of cholesterol induce apoptosis in endothelial and smooth muscle cells in the vasculature. Caspase 9 plays a direct role in modulating apoptosis [48, 52]. ILGFR1 mediates the activity of the insulin like growth factor, and this has a number of mitogenic effects in the cardiovascular system, some of which are linked to hypercholesterolemic conditions [1, 19]. High levels of serum cholesterol, particularly low density lipoproteins (LDL), activate the different components of the arterial wall. The vascular smooth muscle cells, in response to high LDL levels, undergo proliferation, differentiation, as well as apoptosis as a part of atherosclerosis development [1, 2, 47]. One of the mechanisms by which high LDL levels are believed to activate vascular smooth muscles is through the modulation of insulin like growth factor receptor expression. The combination of age and cholesterol lends individuals to an even higher risk for atherosclerosis. Two genes in the ApoE.Age.Diet risk group, Interleukin 1 receptor antagonist (IL1RN) and Clusterin, participate in all aspects of damage for both diet and aging—inflammation, apoptosis and oxidative stress. IL1RN is a naturally occurring antagonist for the inflammatory cytokine interleukin 1 (IL1). Animals deficient for IL1RN develop significantly higher levels of serum cholesterol and atherosclerosis relative to animals with normal ILRN levels [14, 30]. These animals also develop uncontrolled inflammation, particularly within the vascular tree. These animals are also particularly susceptible to the apoptotic effects of elevated cholesterol. In humans, elderly individuals have been shown to have significantly higher levels of circulating IL1RN relative to younger individuals [46]. The overproduction of IL1RN is believed to counter the proinflammatory and proapoptotic effects of age. Clusterin is a glycoprotein that is widely distributed throughout the body [8]. It has been shown to reduce oxidative stress in a number of different biological contexts, such as cigarette smoke exposure and animal models of autoimmune disease, and also to play a role in modulating inflammation [8, 49].

These are a few examples of a larger set of genes showing significant association with the key risk groups and for which there are clear and defensible biological mechanisms by which they may be implicated in the atherosclerotic disease process, though to our knowledge no direct evidence has been earlier identified in the literature.

**7. From mouse to human: Perspective.** From our prior studies of expression from RNA in the inner walls of human aortic tissue [51], we have expression profiles on $n = 122$ individuals. This prior data set also includes the (albeit rather crude) measures of the extent of disease evident within the aortas in terms of percent area affected with raised lesions and also the SudanIV staining measure, as discussed in Section 2. This data set provides the opportunity for an exploratory analysis to evaluate: (a) the extent to which genes identified in key risk groups in the mice models study show evidence of variation in expression in the human samples; (b) the nature of patterns of variation and covariation of the disease related mice genes when explored in the human data; and (c) the extent to which disease risk signatures developed in the mice study relate—if at all—to disease measures in the human samples.

The first step is to link gene probesets across the two microarray platforms, matching genes on the mouse U74av2 array to orthologous sequences on the human U95av2 array. With the current Affymetrix mice and human arrays, this naturally involves cases in which no multiple human gene probe sets are linked to a single mouse probe set, other cases in which multiple human probes sets are matched to a single mouse gene, and cases in which one human gene/probe set is matched to multiple mouse probe sets. In the first case, when no human probe set exists on the Affymetrix U95av2 chip, we simply ignore that mouse gene. In the second case, multiple human probe sets identified as orthologous to the gene represented are evaluated for sequence similarity and a single human gene/probe set selected based on sequence match, the others being ignored. In the third case, we focus only on the unique human genes/probe sets identified. The analysis of the mice experiments described above was in fact performed following such a mapping, this preliminary matching playing a role in defining the number of mice array sequences as mooted earlier. The human U95av2 array has 12588 primary probesets and an additional 67 "AFFX" control probes; the mouse U74av2 array has 12422 primary probesets and an additional 66 control probes. When mapped, the consensus groups involve 7381 gene probesets and 41 common AFFX controls. The former was reduced to 5328 as mentioned, and the data on the latter define assay artifact control covariates.

**8. Mice model risk signatures evaluated in human samples.** Following the mapping of a set of mouse genes $Q$ to the corresponding human genes, we can evaluate the *signature* defined by the metagene $a_{Q,1}$ of the mice gene subset (Section 5.4) against the human data. If $X_Q$ now represents the human data matrix on this gene subset (genes as rows, samples as columns), then the extrapolated signature across human samples is simply $a'_{Q,1} X_Q$. Following the use of metagene signatures in the mice model analysis, we apply this construction not to the raw human data $X_Q$ but instead to human data corrected for potential artifacts. The sparse factor model analysis described in the following section incorporates assay artifact correction terms completely parallel to those included in the mouse model analysis, providing for estimation of the underlying patterns of covariation in the human expression data in the context of a model able to correct—to some degree—for experimentally induced noise and artifactual biases.

Figure 6 displays the projected signatures for four selected groups: the three key gene-environmental risk groups and the older, Western diet-fed, Wild Type mice for contrast. The mice data appear first in each frame. Signatures plotted in the first three frames are those of the risk groups as in Figure 5; in contrast to these, the Age.Diet mice signature in the fourth frame (lower right) indicates very limited discrimination between the older, Western diet fed Wild Type mice and the rest. Mice signatures are followed by the projections onto human samples. The red human cases are positive for raised lesions, while blue cases have no visible evidence of lesions.

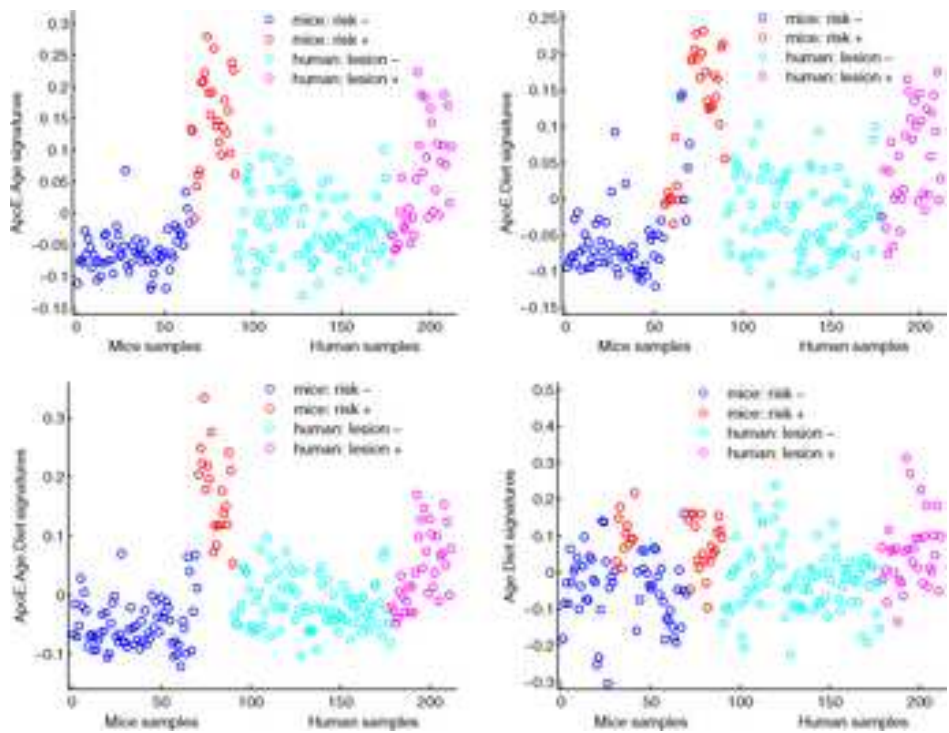

Fig. 6. *Selected expression signatures of risk-related gene groups evaluated in mice samples and projected onto human samples (see also Figure 2 of the supplementary material).*

As discussed earlier, the raised lesion measure is a relatively crude indicator of disease state and progression. Among other questions of its precision as a disease phenotype, it is clear that some individuals with advanced disease may show no evidence of lesioning on the surface of the inner walls of the aorta at all, lesions may be missed in inspection due to selection of tissue regions to examine, lesioning may be evident very locally within a small area of the aorta, and gene expression is defined on mRNA extracted from one isolated section of aortic tissue. Nevertheless, as the only available disease state measure, it is of interest to note that the projected expression signatures of the three key risk groups (ApoeE.Age, ApoE.Diet, ApoE.Age.Diet) show evidence of some discrimination; from Figure 6, roughly half of the "diseased" human cases have high signatures corresponding to these three diseased states in the mice. Though purely exploratory, this does suggest that disease-risk related expression markers can be extrapolated with value from animal to human contexts, and that some degree of variation across samples due to advancing disease is evident in the projection, in spite of the potential shortcomings of the data and the cross-species extrapolation.

## 9. Sparse factor analysis of human gene expression data.

9.1. *Goals and model framework.* Among the questions arising in mapping experimentally derived signatures to observational data are those of just how unrealistic the underlying, experimentally induced changes in gene expression are relative to patterns of normal biological variation. This is a question faced in the current study as in other such studies of cancer related genes [4, 21], for example. Some of the genes perturbed in the experimental context may show little or limited variation at all in normal biological samples, and the extent of expression change in key experimental signature genes is likely to be dramatic compared to normal fluctuations. Further, the complexity of interactions of relevant pathways can be expected to induce more complex patterns of association that are evident in the human observational samples. To explore these questions, we use latent factor models to capture aspects of expression covariation in the human data set, focused on sets of genes that include those identified in the experimental risk signatures. This is a direct application of a general latent factor regression model approach that simply extends the sparse multivariate regression model to incorporate regressors that are uncertain [9]—latent factors reflecting the complexity of purely observational associations among genes evidenced in the human data set. We can then ask how estimated factors relate to the signatures projected from the mouse experiments. This provides some opportunity to examine concordance of projected signatures with key aspects of "normal" patterns underlying gene expression.

We use the same basic notation, now applied to the human gene expression data rather than the mice. On each human sample $i = 1:122$, the $p$-vector of gene expression responses $x_i$ is modeled as

$$(9.1) \qquad x_i = A\lambda_i + Bh_i + \varepsilon_i.$$

Now $B$ is the design matrix that includes a column of 1's followed by the values of the first five assay artifact factors computed on the human array housekeeping genes, so that $\beta_g$ represents the gene-specific intercept and artifact regression parameters. This is precisely as in the mouse model analysis, though without the design components, of course. The additional factor model structure involves the $k$-vector of latent factors $\lambda_i$ and the corresponding sparse factor loadings matrix $A$. The structure and set-up is precisely as described in [9] and also [35]. At a mathematical level the model is just an extension of the multivariate sparse regression in which the $\lambda_i$ vectors are themselves now uncertain. In [9] we describe the use of nonparametric model components for the distribution of the latent factors, a model that permits flexibility in evaluation of potentially complicated patterns of interdependencies among genes in $x_i$ while adapting to what may be a quite non-Gaussian structure. The model involves Dirichlet process priors for the latent factor distribution, and also has the feature of cutting back to Gaussianity if the data suggests that is reasonable.

The analysis parallels that of the sparse anova regression model in that it delivers posterior probabilities $\pi_{g,j}^*$ for all genes $g$ in the model and all latent factors $j = 1\!:\!k$, together with the sparse regression components involving the assay artifact controls. The analysis was initiated with a set of 63 initial genes from the three key mice risk sets. Aiming to select a small initial core set of genes relevant to atherosclerosis risk as defined by the mice data analysis, we selected subsets at the genes for which $\pi_{g,j}^* > 0.99$ in the mice study for these three risk groups $j = 6, 8, 13$, restricting to at most 25 genes from each of the three sets. This yielded a total of 63 unique genes to initiate the evolutionary factor model search analysis [9]. Briefly, this analysis iteratively refits the latent factor regression model, at each iterate based on a "current" set of genes beginning with these 63. Within each step the model is revised to expand the number $k$ of latent factors fitted and that each show association with more than a few genes. This is followed by an approximate computation of the *predictive* probability of gene-factor loadings for all genes not currently included; that is, for all genes $g$ not currently in the model, probabilities analogous to $\pi_{g,j}^*$ for all factors $j$ in the current model. This provides a ranking of nonincluded genes according to how strongly they appear to associate with each of the latent factors. The model can then be extended by including some such genes, assuming subsets show relatively high gene-factor associations. Once the gene set is extended, the factor model is refitted, including possible increase in the number of latent factors. Here we ran this analysis to allow genes to be included only if their maximum predicted probability of inclusion exceeded 0.75, and a factor to be added to increase the dimension of $\lambda_i$ only when at least five genes currently in the model showed $\pi_{g,j}^* > 0.75$. We also restricted the evolutionary expansion of the model to terminate with at most 150 genes in total and the terminal model has $k = 10$ latent factors on 150 genes. At each stage, the MCMC used burn-in of 2000 iterates followed by a Monte Carlo sample of size 8000.

The key concept underlying the evolutionary analysis is to enrich the initial set of "risk related genes" with genes that share latent factor-based associations within the human observational data set; the final model is then an enriched representation of "normal" variation among putatively risk related genes. Full details of the modeling and computational strategy appear in the above reference, and the MCMC analysis of this class of latent factor regression is implemented and exemplified in the BFRM software.

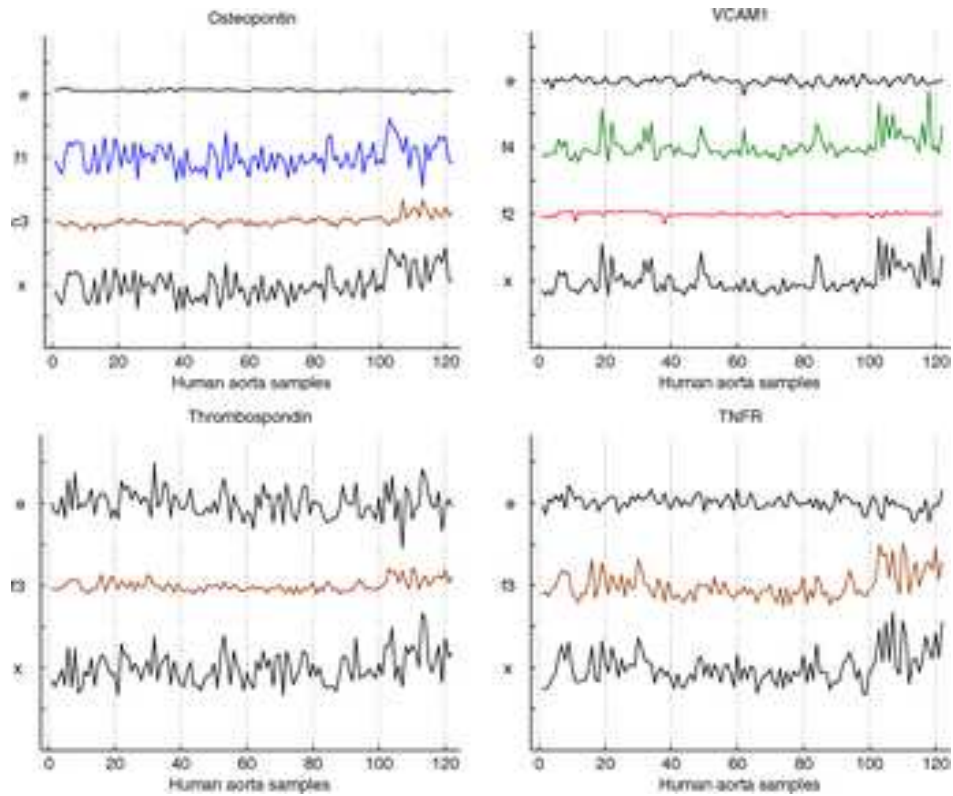

Fig. 7. *Expression decompositions across human aorta samples of the four risk-related genes illustrated in Figure 3. In each frame, the plots over time are expression data $x$, fitted model components from the subset of latent factors and assay artifact controls showing significant $(\pi_{g,j}^* > 0.95)$ contributions (if any) for that gene, and residuals. Expression is effectively the direct sum of the components and residuals plotted, and within each frame the data and components are plotted on the same vertical scales for comparison. Labels $f1, f2, \ldots$ represent fitted latent factor effects for factors 1, 2, and so forth, $c3$ the fitted effects of the third assay artifact correction factor, and $e$ represents residuals.*

9.2. *Aspects of risk-related factor structure in human data.* Figure 7 displays data for the four example "risk-related" genes identified in the mice data analysis (Figure 3). As in the mice analysis figure, these represent decompositions of expression across samples, with each gene decomposed into significant contributions from the model factors or regressors. The samples are ordered as earlier so that the last 33 samples are those showing evidence of more advanced disease in terms of raised lesions.

It appears that each of these genes does indeed tend to show higher levels of expression in later samples, though with much variation and no clear or really obvious step or ramp-up of levels among the "diseased" cases. Two of these genes (Thrombospondin and TNFR) are, among others, significantly associated with latent factor 3; the posterior mean of factor 3 across samples shows a clearer pattern of higher values among later samples, though again with volatility. Thus, factor 3 seems to be an inherent pattern in the observational data set that relates, to a degree, to the increased disease risk measured by lesions, and reflects a key contributing pattern to the overall expression of these known risk marker genes. Among the 150 genes in the analysis, 37 show significant association ($\pi^*_{g,j} > 0.95$) with factor 3, including a number from the initial risk set but also additional genes that were identified as part of the evolutionary analysis. This points to the potential for this form of analysis to take a step-ahead in identifying additional gene candidates for further study in connection with disease risk processes, and some of these genes not previously identified as risk related are worth investigating further. Factor 4 also appears to relate to risk, as evidenced in the plot for VCAM1. Factor 4 appears significantly loaded on a smaller, biologically very cohesive set of genes that include representative probesets for VCAM1 and others, and as apparent in the decomposition of VCAM1 in the figure, shows a (weak) relationship with the more diseased individuals. Thus, the factor analysis has identified two distinct patterns of common association in the normal variation of expression profiles, each related to risk and potentially reflecting underlying substructure in disease related pathway activity. Finally, Figure 8 shows the concordance between the estimated factor 3 and the projection from the mice data analysis of the metagene signature corresponding to the $j = 6$ (ApoE-/-, Age) risk factor group. The remarkable concordance indicates that the overall pattern of natural variation in expression that factor 3 captures and reflects is inherently consistent with that observed in the controlled and artificial experimental context. This gives further credence to the view that the gene subsets underlying these factors and signatures are indeed likely to be both predictive of disease states in human contexts and suggestive of directions to investigate for improved biological understanding.

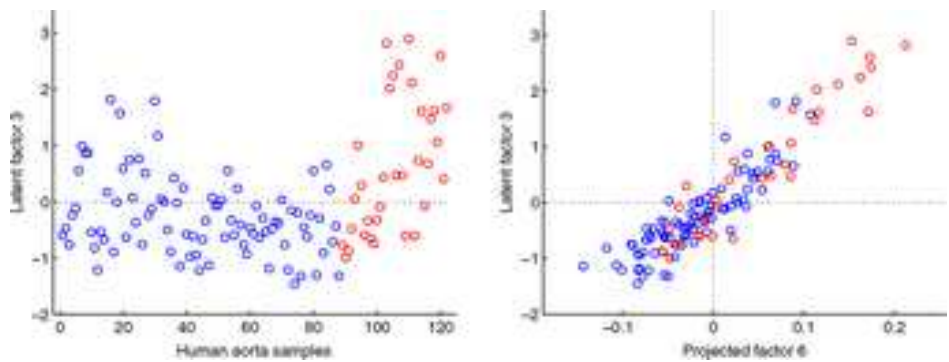

FIG. 8. Left frame: *The posterior mean of latent factor $f3$ in the human aorta analysis, plotted across samples. The red cases are those for which there is evidence of raised lesioning within the inner wall of the aorta, that is, the more advanced disease cases, with lesion index increasing with sample index number (see Figure 2 of the supplementary material).* Right frame: *The same estimated factor 3 scatter-plotted against the projected risk signature from the $j = 6$ (ApoE-/-, Age) risk factor group from the mice data analysis.*

**10. Concluding comments.** The analysis of the experimental mice data generates robust patterns of expression that are influenced by the interactions of the effects of key, known risk factors; these patterns can be characterized by metagene signatures of the effects and interactions of the gene-environmental factors. When mapped to human expression data, with all of the issues of data comparability, cross-species and technological issues that raises, these patterns relate to the key current measure of extent of atherosclerotic disease burden in observational human aorta samples. The groups of genes identified clearly link to much of what is known about the disease process and its development with age and as response to diet, while also isolating additional genes that are potentially important modifiers of atherosclerotic risk in the setting of specific risk factors. This application is an example of an approach to dissecting gene-environment interactions on a genomic scale in the context of a critical, complex human disease, and one that is being used in other areas including cancer studies.

The study also serves to illustrate the utility of sparsity modeling in multivariate, high-dimensional anova, regression and latent factor models. Although the statistical framework and methodology used for these applications are generic and applicable in other fields, a major motivation in their development has been biological pathway and translational studies using gene expression data, and other forms of high-throughput molecular data in both designed experimental contexts and observational studies. As high-throughput arrays and related technologies become desk-top commodities, we can expect to see a major increase in the scale and complexity of designed experiments on multiple factors generating high-dimensional responses. Factorial designs will become *de rigeur* within molecular and genome biology in the way they were in the early 20th century in agricultural research, and the need for relevant statistical analysis tools will be evermore central. Sparsity and shrinkage modeling are fundamental and, though the specific forms of models used here are just examples, the framework is clearly extensible to other similar contexts. Coupled with that, sparse latent factor models for observational data sets provide complementary approaches to deconvolution and attribution of complex, interacting patterns of association in data that may reflect inherent, underlying biological processes and pathway substructure. As in cancer applications [9, 35], these methods can play roles in studies to elucidate gene-environment interactions in critical and challenging biomedical contexts such as atherosclerosis.

**Acknowledgments.** The authors are grateful to an Editor of AOAS for constructive comments on this paper. Any opinions, findings and conclusions or recommendations expressed in this work are those of the authors and do not necessarily reflect the views of the NSF or NIH.

**Supplements, code and data.** The BFRM software, an implementation of MCMC analysis of sparse statistical models for multivariate data with latent factor, regression and ANOVA components, is available for interested researchers. The software, instructions and examples are available at <http://xpress.isds.duke.edu:8080/bfrm/>. Data, parameter input files and output files from the BFRM analyses summarized here are also available, together with some additional tables and figures, as supplementary material at <http://ftp.stat.duke.edu/WorkingPapers/07-05.html> and also linked to the paper at the journal web site.

D. M. Seo
Institute for Genome Sciences and Policy
Duke University
Durham, North Carolina 27710
USA
E-mail: david.seo@duke.edu
URL: www.genome.duke.edu

P. J. Goldschmidt-Clermont
Miller School of Medicine
University of Miami
Coral Gables, Florida 33124
USA
E-mail: pgoldschmidt@med.miami.edu
URL: www.med.miami.edu

M. West
Department of Statistical Science
Duke University
Durham, North Carolina 27708-0251
USA
E-mail: mike@stat.duke.edu
URL: www.stat.duke.edu/~mw